**Metabolomic signature of type 1 diabetes-induced sensory loss and nerve damage in diabetic neuropathy.**

Daniel Rangel Rojas[1], Rohini Kuner[1], Nitin Agarwal[1]*

[1]Institute of Pharmacology, Heidelberg University, Im Neuenheimer Feld 366, D-69120 Heidelberg, Germany.

*Address correspondence to N.A, Institute of Pharmacology, Heidelberg University, Im Neuenheimer Feld 366, D-69120 Heidelberg, Germany. E mail: nitin.agarwal@pharma.uni-heidelberg.de

## ABSTRACT

**Objective:** Diabetic-induced peripheral neuropathy (DPN) is a highly complex and frequent diabetic late complication, which is manifested by prolonged hyperglycemia. However, the molecular mechanisms underlying the pathophysiology of nerve damage and sensory loss remain largely unclear. Recently, alteration in metabolic flux have gained attention a basis for organ damage in diabetes; however, peripheral sensory neurons have not been adequately analyzed with respect to metabolic dysfunction. In the present study, we attempted to delineate the sequence of event occurring in alteration of metabolic pathways in relation to nerve damage and sensory loss.

**Methods:** C57Bl6/j wild type mice were analyzed longitudinally up to 22 weeks (wks) in the streptozotocin (STZ) model of type1 diabetes. The progression of DPN was investigated by behavioral measurements of sensitivity to thermal and mechanical stimuli and quantitative morphological assessment of intraepidermal nerve fiber density. We employed a mass spectrometry-based screen to address alterations in levels of metabolites in peripheral

**Abbreviations:** ROS - reactive oxygen species; STZ -Streptozotocin; DPN - Diabetic peripheral neuropathy; SN- sciatic nerve; SNS – sensory neuron-specific; MS – Mass spectrometry sciatic nerve and amino acids in serum over several months post-STZ administration to elucidate metabolic dysfunction longitudinally in relation to sensory dysfunction.

**Results:** Although hyperglycemia and body weight changes occurred early, sensory loss and reduced intraepithelial branching of nociceptive nerves was only evident at 22 wks post-STZ. The longitudinal metabolites screen in peripheral nerves demonstrated that compared to buffer-injected age-matched control mice, mice at 12 wks and 22 wks post-STZ showed an early impairment the tricaoboxylic acid (TCA cycle), which is the main pathway of carbohydrate metabolism leading to energy generation. We found that levels of citric acid, ketoglutaric acid (2 KG), succinic acid, fumaric acid and malic acid were observed to be significantly reduced in sciatic nerve at 22 wks post-STZ. In addition, we also found the increase in levels of sorbitol and L-Lactate in peripheral nerve from 12 wks post-STZ injection. Amino acid screen in serum showed that the amino acids Valine (Val), Isoleucine (Ile) and Leucine (Leu), grouped together as BCAA, increased more than 2-fold from 12 wks post-STZ. Similarly, the levels of Tyrosine (Tyr), Asparagine (Asn), Serine (Ser), Histidine (His), Alanine (Ala), and Proline (Pro) showed progressive increase with progression of diabetes.

**Conclusion:** Our results indicate that the impaired TCA cycle metabolites in peripheral nerve is the primary cause of shunting metabolic substrate to compensatory pathways which leads to mitochondrial dysfunction and nerve damage.

**Introduction**

Diabetes is a chronic metabolic disease marked by hyperglycemia as a result of dysfunction in insulin secretion and/or action. Data from the World Health Organization shows that global prevalence of diabetes among adult is 8.5% [1]. Diabetes-induced peripheral neuropathy (DPN) is a highly complex and prevalent diabetic complication observed in 50 % of diabetic patients. DPN is characterized by progressive loss of peripheral nerve axons resulting in pain, loss of sensation, and eventually a leading cause of lower extremity amputation [2, 3]. Despite a long history of research on delineating the pathophysiology of the disease we are now beginning to understand molecular mechanisms underlying DPN.

Hyperglycemia-induced oxidative stress is postulated to be a primary driver in diabetic complication and organ dysfunction [4]. Several studies showed that hyperglycemia causes to mitochondrial dysfunction leading to overproduction of superoxide and free radicals [5]. Elevated ROS levels trigger alterations in transcriptional factors function, release of inflammatory cytokines and chemokines. Other cellular consequences include the release of cytochrome c, activation of caspase 3, activation of an endonuclease-G (Endo-G), altered biogenesis and cell death [6]. Hyperglycemia-induced ROS generation is linked to both enzymatic and nonenzymatic pathways. The enzymatic pathways include nicotinamide adenine dinucleotide phosphate oxidase (NADPH oxidase) and uncoupling of nitric oxide synthase (NOS). The nonenzymatic pathways include aberrant functionality of complex 1 and complex III of Electron Transport Chain (ETS). The metabolic inflexibility of cell to switch between nutrient utilization, in particular between fatty acid and glucose oxidation is postulated to be the primary reason for ROS formation. Emerging evidence suggests that alterations in glycolytic flux, signaling molecules and mitochondrial enzymes contribute to tissue-specific adaptations in fuel utilization that are associated with tissue dysfunction [7].

Rodent and human studies results showed the down regulation of expression and activity of electron chain transport proteins in diabetic heart [8, 9] and skeletal muscles of patients with type 2 diabetes [10, 11]. Similarly downregulation of mitochondrial enzymes in DRG and nerves have been observed after established DPN in genetically modified diabetic mice models which do not replicate the clinical observation associated with disease progression.[12]. Moreover, the loss of peripheral nerve fibers in the epidermal layer of skin is considered to be a clinical diagnostic tool for DPN detection. Although these studies showed alterations in activity and expression of metabolic pathways enzymes in DRG, peripheral nerves and nerve fiber in skin but what remains unanswered is whether these events occurring after established DPN or they are the reasons for onset and development of DPN. However, to date, no attempts were made to investigate the progressive change happening in metabolic pathways before and during the development of DPN. Branched-Chain Amino Acid (BCAA), glucogenic and ketogenic amino acids have reported to have critical role in metabolic health. Cross section-studies in type 2 diabetes patients and in rodents have reported elevated levels of BCAA in serum contribute to the development of insulin resistance. Lack of longitudinal studies limit the association of amino acid to the progression of DPN. We therefore performed a comprehensive and unbiased amino acid screen to identify the signature tracking the progression of DPN in the plasma of STZ-treated mice as compared to control buffer-injected mice.

In the present study, we have systemically analyzed the change in the kinetics of tri-carboxylic cycle (TCA) in peripheral nerve and amino acids in serum by employing mass spectrum based measurement of TCA metabolites and amino acids at different point of time in mouse model of type 1 diabetes. We attempt to establish the link between observed metabolic alterations to pathological symptoms of DPN by performing behavioral analysis and changes in peripheral nerve fiber density in epidermal layer of skin. We report a series of changes in amino acids in nerves of diabetic mice and show that inhibition of glycolytic enzymes is an early event occuring in response to hyperglycemia.

## 2    Material and Methods

### 2.1    Animal experiments

Age-matched 7-8 wks old C57BL6/j male mice were bought from Janvier labs, Europe. Animals were maintained in humidity and temperature controlled environment. Mice were housed in socially stable and well nested individually ventilated cage-rack system (Techniplast, Italy) and had free access to water and food. All experiments were done in accordance with ARRIVE guidelines and approved by Regierungspräsidium Karlsruhe, Germany.

## 2.2 Rodent model of type 1 diabetic model

We employed low dose streptozotocin (STZ)-induced type 1 diabetic model for all the experiments. Multiple low dose STZ (60 mg/kg/d, for 5 consecutive days) intraperitoneal (i.p) administration leads to selective destruction of pancreatic beta cells. It is evident by increased levels of blood glucose. Blood glucose levels were monitored weekly using glucometer (Accu-Chek, Roche Diagnostics) for the entire course of the experiment. Mice only with blood glucose levels >350mg/dl were considered to be diabetic and included in experiments. Mice were analyzed over a period of 8 wks to 22 wks post-STZ injection. Post-STZ mice blood glucose levels were maintained in a range between 350 and 500 mg/dl by sub-cutaneous administeration of insulin.

## 2.3 Tissue extraction

Blood and tissue samples were collected at pre-diabetic and at different time point post-STZ injection. Age-matched citrate buffer injected, non-diabetic mice were used as controls for blood and tissue collection. Mice were anesthetized under 2% isoflurane. The blood samples were drawn from the venous sinus using the retro-orbital bleeding method. The blood samples were collected from pre-diabetic, 8, 12, and 22 weeks (wks) post-STZ injection in EDTA-treated tubes and plasma was separated by centrifugation. Sciatic nerve tissue was extracted from pre-diabetic, 12 and 22 wks post-STZ injected diabetic mice. The separated plasma and extracted sciatic nerve was snap freezed using liquid nitrogen and stored at -80°C until analysis.

For paw punches biopsies, the mice at different time point post-STZ were perfused with 4% paraformaldehyde (PFA). The punch biopsies of the plantar skin of the hind paws were prepared and post-fixed in 4% PFA for 24 hours (hrs) at 4°C. Tissue was incubated overnight in 30% sucrose and 16 μm cryosections were made for immunohistochemical analysis.

## 2.4 Immunohistochemistry

To evaluate the degeneration of nerve fibers in the epidermal layer of diabetic mice, we performed immunohistochemistry using anti-CGRP antibody [13]. Immunostaining was performed on punch biopsies of the plantar skin samples extracted from diabetic and non-diabetic C57BL6/j mice prior to and different time points post-STZ as described before. Briefly, cryosections were permeabilized in 0.5% PBST, washed and blocked with 7% Horse serum (HS). Sections were incubated overnight at 4°C with anti-CGRP (1:1000, sigma) primary antibody in 7% HS in PBS. Subsequently, sections were washed and incubated with Alexa Fluor-594-conjugated secondary antibody. Sections were washed and mounted in Mowiol. Fluorescence images were obtained using a laser-scanning spectral confocal microscope and maximal projections were created using Leica SP8 software (Leica TCS SP8 AOBS, Bensheim, Germany). The acquired images were analyzed to quantify the fluorescence intensity from the epidermal area using Image J software.

## 2.5 Behavioral measurements

All behavioral experiments were approved by Regierungspräsidium Karlsruhe, Germany. Mice were acclimatized to the experimental setup twice a day for 3 days. The animals were randomized and the experimenter was blinded for the identity of treatment given to mice. Age-matched citrate buffer injected control mice were used for all behavioural experiments. Thermal sensitivity was measured using Hargreaves apparatus. Briefly, the paw withdrawal latency on application of infrared (IR) heat source on the plantar surface of hindpaw was recorded with a cut off of 25 sec. Two consecutive heat applications on the same mouse were separated with time interval of 5 minutes. Mechaincal sensitivity was analysed using Von Frey mono filaments. Mice were placed on an elevated grid and von Frey monofilaments with specific forece were applied to the plantar surface of hindpaw. von Frey filament of 0.16, 0.4, 0.6, 1.0, 1.4, 2.0 and 4.0 g were tested to determine the mechinal sensitivity. The force range was chossen such that the mice response from no response to 100 % response. Each monofilament was applied for 5 times at time interval of 10 min on the plantar hindpaw. 40% response frequency was calculated as "thresholds" at basal, 12 and 22 wks post-STZ injection.

### 2.6 Determination of amino acid levels

Amino acids from mice serum were quantified after specific labeling with the fluorescence dye AccQ-Tag$^{TM}$ (Waters) according to the manufacturer's protocol. The resulting derivatives were separated by reversed phase chromatography on an Acquity BEH C18 column (150 mm x 2.1 mm, 1.7 µm, Waters) connected to an Acquity H-class UPLC system and quantified by fluorescence detection (Acquity FLR detector, Waters, excitation: 250 nm, emission: 395 nm) using ultrapure standards (Sigma). The column was heated to 42 °C and equilibrated with 5 column volumes of buffer A (140 mM sodium acetate pH 6.3, 7 mM triethanolamine) at a flow rate of 0.45 ml/min. Baseline separation of amino acid derivates was achieved by increasing the concentration of acetonitrile (B) in buffer A as follows: 1 min 8% B, 16 min 18% B, 23 min 40% B, 26.3 min 80% B, hold for 5 min, and return to 8% B in 3 min. Data acquisition and processing were performed with the Empower3 software suite (Waters).

### 2.7 Determination of metabolites by gas chromatography/mass spectrometry from sciatic nerve tissue

#### 2.7.1 Extraction

Frozen material was extracted in 180 µl of 100% MeOH for 15 min. at 70°C with vigorous shaking. 5 µl Ribitol (0.2 mg/ml) was added as internal standard to each sample. It was followed by addition of 100 µl chloroform to each sample. All samples were shaken at 37°C for 5 min. Subsequently, 200 µl of water was added to each sample and centrifuged at

11,000 g for 10 min. Three hundred microliters of the polar (upper) phase were transferred to a fresh tube and dried in a speed-vac (Eppendorf vacuum concentrator) without heating and used for derivatization.

### 2.7.2 Derivatization (Methoximation and Silylation)

Pellets were re-dissolved in 20 µl methoximation reagent containing 20 mg/ml methoxyamine hydrochloride (Sigma 226904) in pyridine (Sigma 270970) and incubated for 2 hrs at 37°C with shaking. For silylation, 32.2 µl N-Methyl-N-(trimethylsilyl)-trifluoroacetamide (MSTFA; Sigma M7891) and 2.8 µl Alkane Standard Mixture (50 mg/ml $C_{10}$ - $C_{40}$; Fluka 68281) were added to each sample. After incubation for 30 min at 50°C, samples were transferred to glass vials for Gas Chromatography/Mass Spectrometry (GC/MS) analysis.

### 2.7.3 GC/MS analysis

GC/MS-QP2010 Plus (Shimadzu®) fitted with a Zebron ZB 5 MS column (Phenomenex®; 30 m x 0.25 mm x 0.25 µm) was used for GC/MS analysis. The GC was operated at an injection temperature of 230°C and 1 µl sample was injected with split mode (1:10). The GC temperature program started with a 1 min. hold at 40°C followed by a 6°C/min ramp to 210°C, a 20°C/min ramp to 330°C and a bake-out for 5 min at 330°C using Helium as the carrier gas with constant linear velocity. The MS was operated with ion source and interface temperatures of 250°C, a solvent cut time of 8 min and a scan range (m/z) of 40 - 700 with an event time of 0.2 sec. The "GCMS solution" software (Shimadzu®) was used for data processing.

### 2.8 Statistical analysis

All the data were calculated and are presented as mean ± SEM. ANOVA for repeated measures followed by Bonferroni's test for multiple comparisons was employed to determine statistically significant differences. Changes with $p \leq 0.05$ were considered to be significant.

## 3   Results

### 3.1   Induction of STZ-induced DPN in mice

This study investigates alterations in amino acid profile in serum and the metabolic changes occurring in sciatic nerve over the course of progression of DPN. We employed the low dose STZ model of type 1 diabetes, which does not involve direct neurotoxic effects. [14]. It is characterized by lymphocytic infiltration of pancreatic islets, leading to cell death, subsequently resulting in insulin deficiency and hyperglycemia [15]. We performed a long-term study in STZ-injected and age-matched citrate buffer-injected control mice to investigate the role of hyperglycemia induced metabolic changes in the onset and development of DPN. Persistence of diabetes was examined by measuring blood glucose levels and we had blood glucose levels to maintain all animals at uniform levels of hyperglycemia within the cohort. STZ-injected mice incessant showed blood glucose levels of > 400 mg/dl from 2 wks onwards post-STZ injection over the entire course of the disease (Figure **1A**, $p<0.05$, ANOVA). STZ-injected did not suffer any weight loss over basal but also did not gain any weight as age progressed, unlike buffer-injected age-matched controls at all time points (Figure **1B**, $p<0.05$, ANOVA).

We investigated STZ-injected diabetic and age-matched citrate-buffer injected mice in the behavioral model of evoked pain at basal, 12 and 22 wks post-STZ injection to evaluate the course of DPN. Although behavioral changes have been reported [16], it was important to assess metabolic changes and sensory abnormalities in the same cohort of mice to derive meaningful relationships. The time point chosen for analyses is based on rational to address onset and development of DPN. STZ-treated mice showed no change in response threshold as compared to control group on plantar application of von Frey monofilaments at 12 wks post-STZ injection (Figure **1C**). Similarly, in Hargreaves test, we did not find any significant difference between STZ-injected and control groups in withdrawal frequency at 12 wks post-STZ injection (Figure **1D**). However, at 22 wks post-STZ we found that STZ-injected group of mice developed mechanical and thermal hypoalgesia which demonstrate the pathophysiology of established DPN (Figure **1C** and **D**, * p<0.05 Two-way ANOVA). The control group of mice did not show any sign of neuropathy.

Assessment of Intra-Epidermal Nerve Fibers Density (IENFD) in skin biopsies is an important approach to appraise progression of the DPN, which is also used in the clinical diagnostic context [17]. We attempt to correlate the IENFD with progression of diabetes in STZ treated mice. We performed immunostaining using Calcitonin Gene-Related Peptide (CGRP) on skin sections to label nociceptive nerve endings which are the main class of afferents in the epidermal zone. We found no significant changes in IENFD at 12 wks post-STZ but strikingly at 22 wks post-STZ, IENFD was significantly reduced (figure 1E, F, *p<0.05, ANOVA). The reduced IENFD and behavioral changes at 22 wks post-STZ treatment fitted the sensory loss observed previously in STZ-treated mice [18].

## 3.2  Alteration in amino acid level with onset and progression of DPN.

We performed a longitudinal study to identify the amino acid signature in serum to track the progression of DPN in STZ-treated mice. Citrate buffer-injected control mice do not showed any significant changes in the levels of BCAA, glucogenic and ketogenic amino acids at 12 wks and 22 wks after injection. We found that the amino acids Valine (Val), Isoleucine (Ile) and Leucine (Leu), grouped together as BCAA [19], increased more than 2-fold from 12 wks post-STZ (Figure **2A**, * p<0.05 as compared to basal, ANOVA). Leucine showed nearly 3-fold increase in STZ-treated mice over age-matched citrate buffer injected control mice.at 22 wks. Several gluconeogenic and ketogenic amino acids were unchanged at 12 or 22 wks in STZ-treated mice as compared to controls. For example Aspartate (Asp), Glutamate (Glu), Glutamine (Gln), Glycine (Gly), Methionine (Met), Threonine (Thr) and Phenylalanine (Phe) remained unchanged at 12 ad 22 wks post-STZ injection, whereas Tyrosine (Tyr), Asparagine (Asn), Serine (Ser), Histidine (His), Alanine (Ala), Proline (Pro) (figure **2B**, **C** and **D**, *<0.05, ANOVA) showed progressive increase in plasma at 12 wks post-STZ injection. Our results thus indicate diverse changes in plasma levels of BCAA and some gluconeogenic and ketogenic amino acids building up over time as diabetic complications set in.

### 3.3    Reduced TCA cycle intermediates in sciatic nerve of diabetic mice

The progressive impact of STZ-induced hyperglycemia on the kinetics of glycolytic and TCA cycle has not been described in nerves so far. Therefore, we performed targeted LC/MS-MS-based metabolomics screen. We measured the levels of glycolytic and TCA cycle intermediates in sciatic nerve of mice at basal state, 12 wks and at 22 wks post-STZ injection. The levels of citric acid, ketoglutaric acid (2 KG), succinic acid, fumaric acid and malic acid were observed to be significantly reduced in sciatic nerve at 22 wks post-STZ as compared to age-matched citrate buffer-injected control mice (Figure **3A**, *$p<0.05$, ANOVA). There was a trend for reduction of some of the TCA metabolites at 12 wks, but changes only became significantly evident at 22 wks, i.e. at the stage when sensory loss and nerve damage are clearly present. Progressive reduction in TCA metabolites results in metabolic shift to alternative pathways to meet the energy requirement. We found a continuous increase in levels of sorbitol in the sciatic nerve biopsies (Fig **3B**, *$<0.05$, ANOVA) at 12 wks at 22 wks in STZ-treated mice. Gradual increase in levels of sorbitol and diminishing kinetic of TCA cycle with prolonged diabetic state indicate that there is a shift of glucose metabolism from glycolytic pathway to polyol pathway which is linked to tissue dysfunction. Further, we found the elevated levels of L-lactate (Fig **3C**, *$<0.05$, ANOVA) at 22 wks post-STZ injection.

### 4    Discussion

DPN is the major reason for morbidity among diabetic patients [20]. The pathogenic triggers leading to onset and progression of DPN remain unidentified. As the development of DPN is not a static process, there is a continuous need to identify the markers, which then track specific and temporal alteration in onset and progression of DPN. We attempt to identify the metabolic changes occurring in the peripheral nerve tissue and in serum prior to the development of DPN. The main findings revealed by this study are: (i) the switching of metabolic flux from oxidative pathway to alternative pathways in peripheral nerve is an early event occurring when no detectable symptoms of nerve dysfunction and sensory loss are present, (ii) the alteration in metabolic pathways in sciatic nerve are accompanied by increased levels of BCAA in serum, which may be a potential prognostic marker to detect diabetic complications.

Several studies in diabetic rat and in genetically modified-mice repeatedly reported metabolic pathways dysfunction in both type 1 and 2 diabetes models [21, 22, 23]. Oxidative stress, lipids and proteins oxidations, inhibition of metabolic enzymes, increased BCAA [24], downregulation of mitochondrial enzymes in DRG and nerve have been observed after established DPN [23]. In this study we attempt to understand the link between neuropathic symptoms with metabolic dysfunction. We demonstrated that at 12 wks post-STZ injection there was no detectable nociceptive dysfunction. It is further validated by no detectable changes in nerve fibers density in the epidermal layer of skin. In contrast, at similar state we detect reduced levels of citric acid and a trend towards impairment of TCA cycle. Aconitase enzyme, which catalyzes the conversion of citrate to isocitrate, is the most sensitive TCA enzyme for ROS-inhibition [25]. Hyperglycemia-induced aconitase inhibition is previously reported rodent model with established DPN [25].

Impairment of the TCA cycle deviate metabolites to compensatory metabolic pathways for energy generation. In nerve tissue, it is evident by elevated levels of sorbitol and L-lactate at similar stage. Inceased levels of sorbitol indicate deviation of glucose from glycolysis/TCA pathway to polyol pathway. Elevated L-lactate levels are an indicator of mitochondrial dysfunction and shift from oxidative phosphorylation to anaerobic metabolic pathways. Our results demonstrate the reduced kinetic of TCA cycle leads to metabolic shift to alternative pathway which with prolonged state results in development of DPN. It is conceivable that these changes lead to further dysfunction of energy metabolism and contribute to the onset and development of DPN. At late stage of 22 wks post STZ, DPN was evident by developed mechanical and thermal hypoalgesia, reduced nerve fibers density and complete sink in the levels of TCA intermediates. Previous studies in sural nerve demonstrated that DPN is also marked by reduced expression of glycolytic enzymes and compromised axonal transport.

[26]. Our results indicate that these hyperglycemia-induced site-specific molecular changes in metabolic pathways are the first line of events, which may cause increased distal oxidative stress and contribute to onset and development of DPN.

In an attempt to develop a noninvasive prognostic marker, we correlated the metabolic changes in peripheral nerve tissue to changes in levels of amino acid in serum. In diabetic patient, it was shown that insulin resistance switches the metabolism of cell to utilize glucogenic and ketogenic amino acid as a source of energy. It results in reduction of these amino acids over time. In contrast the levels of BCAAs were reported to be increased in patients with established diabetic complication. Our longitudinal studies demonstrate a trend for increasing levels of BCAAs as early as 5 wks post-STZ, which were significantly elevated at 12 wks and 22 wks post-STZ injection. The levels of tyrosine were also found to be increased from 12 wks post-STZ which has been recently marked a potential marker for obesity-induced modulation of insulin signaling [27]. Asn, His, Ala and Pro showed significant increase at 12 wks post-STZ indicating the development of diabetic complications

## 5 CONCLUSIONS

In summary, we propose that inhibition of glycolytic enzymes is an initiation event in sciatic nerve occurring at a stage when there are no evident symptoms of DPN. Inhibition of TCA cycle shifts glucose metabolism to alternative pathways, which are marked by early increase in the levels of sorbitol and L-lactate in peripheral nerves post diabetes-induction. These early events are paralleled by changes in BCAA serum levels, which come about very early and may hold some prognostic value. These observations from longitudinal analyses provide insights into delineating the temporal sequence of anomalies and may be supportive in designing and testing novel therapeutic measures to prevent or reverse the progression of DPN.

## 6 AUTHORS CONTRIBUTIONS

NA and RK designed the study; DRR and NA performed and analyzed the experiments; NA and RK wrote the manuscript. All authors approved the final version.

## 7 ACKNOWLEDGMENTS

The authors thank Rose LeFaucheur for secretarial help, Dunja Baumgartl-Ahlert and Hans-Joseph Wrede for technical assistance. This work was supported by a grant from the Deutsche Forschungsgemeinschaft (DFG) in the Collaborative Research Center 1118 (SFB1118 Project B06) to N.A. and R.K. We would like to thank the Metabolomics Core Technology Platform of the Excellence Cluster CellNetworks for support with amino acid and metabolite quantification.

## 8 CONFLICT OF INTEREST

The authors declare that there is no conflict of interest.

**Figure legends:**

**Figure 1:** Course and quantitative morphological analysis of DPN-associated pathology of peripheral nerves in STZ-injected C57Bl6/j mice and citrate buffer-injected age-matched control mice **(A)** Blood glucose levels in C57BL6/j after STZ injection **(B)** Course of body weight in mice after STZ-injection. **(C, D)** Summary of response thresholds to mechanical stimuli (defined as a force eliciting a response of paw withdrawal at least 40% response rate) **(E)** Identification of intra-epidermal nerve fibers via immunostaining for CGRP in the paw skin of STZ or buffer injected mice. The dotted white lines represent the epidermal layer in the skin. **(F)** Quantification of reduction in epidermal nerve fiber density post-STZ injection in STZ-injected and control mice. n= 8 mice per group. Data are represented as mean ± S.E.M. *$p < 0.05$ compared to baseline, # $p < 0.05$ as compared to control group two-way ANOVA for repeated measurements with Bonferroni multiple comparison test. Scale bar represents 30 µm.

**Figure 2:** Quantitative analysis of amino acid levels in serum of STZ-injected and control mice. **(A**) Branched amino acids **(B)** Gluconeogenic and ketogenic **(C)** Gluconeogenic amino acids **(D)** other amino acids levels in serum at basal, 8, 12 and 22 wks post-STZ and citrate buffer-injected mice. Fold changes of amino acid in serum of STZ-injected mice were calculated with age-matched citrate buffer-injected controls. N = 5 mice in each group. *$p < 0.05$ compared to baseline, ANOVA for repeated measurements with Bonferroni multiple comparison tests.

**Figure 4:** Massspectrometry-based quantitative analysis of citric acid metabolites in sciatic nerve of STZ-injected mice. **(A)** Citric acid metabolites **(B)** sorbitol levels **(C)** L-lactate levels in sciatic nerve isolated from STZ- and citrate buffer-injected mice at basal, 12 and 22 wks post injection. The graph represents fold change in metabolite levels over age-matched control mice. N = 5 in mice in each group. *$p < 0.05$ compared to baseline, ANOVA for repeated measurements with Bonferroni multiple comparison test.

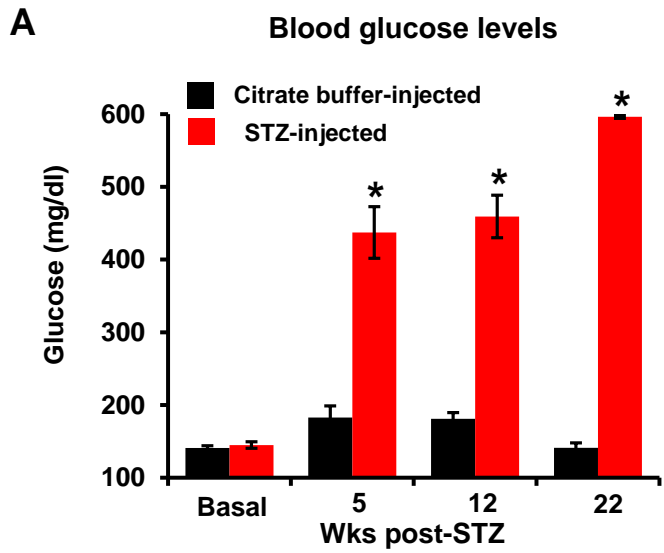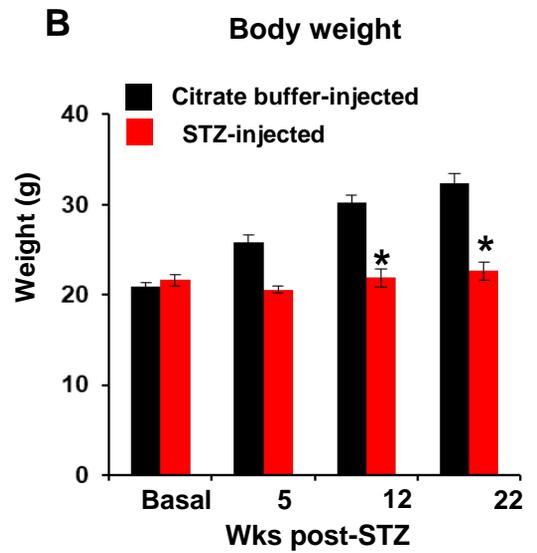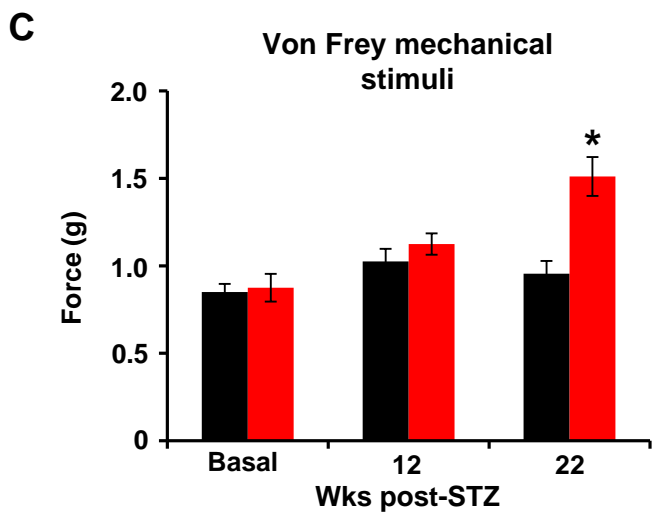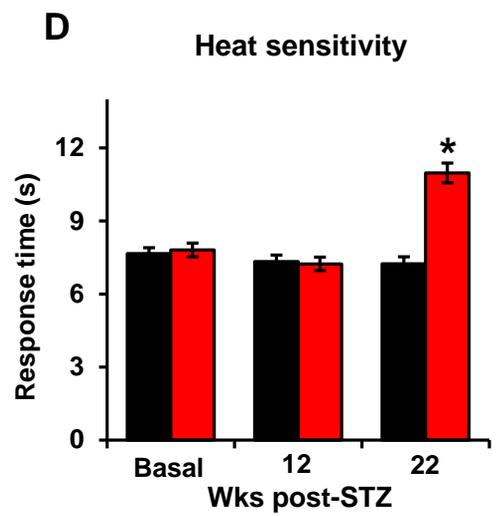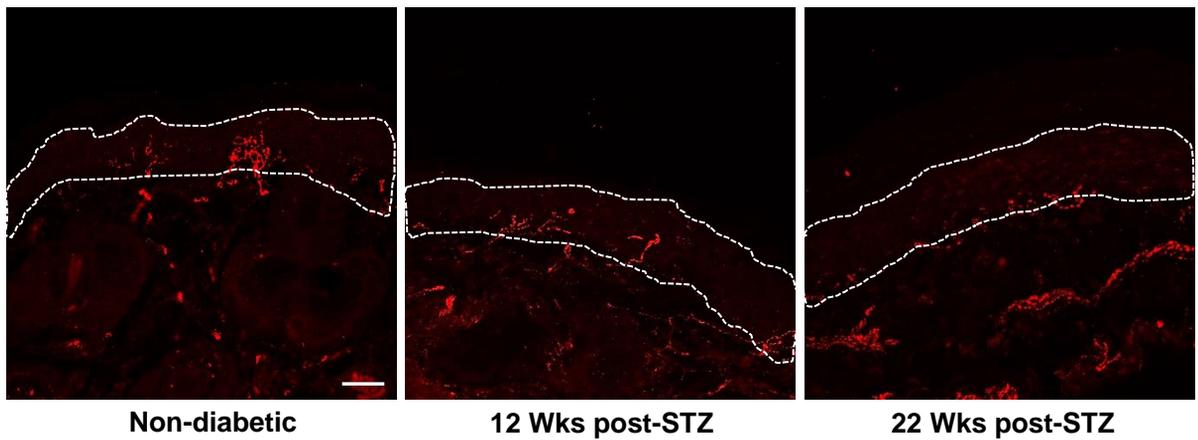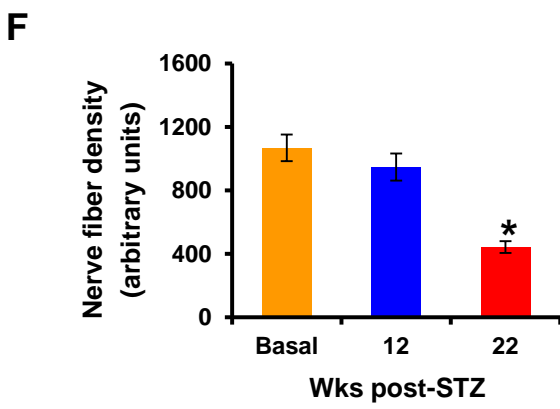

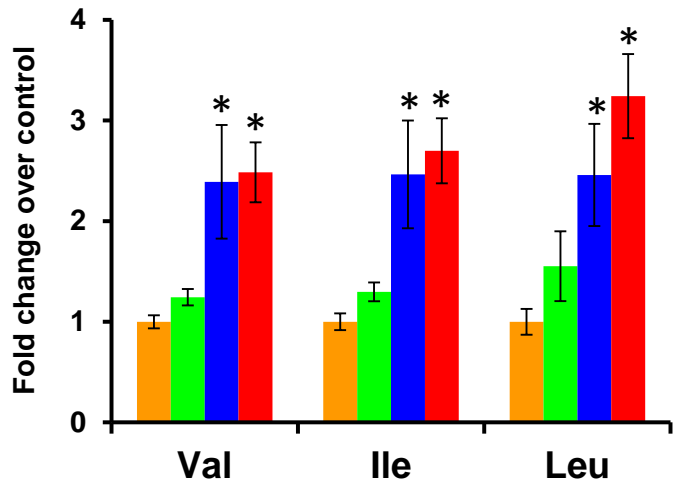
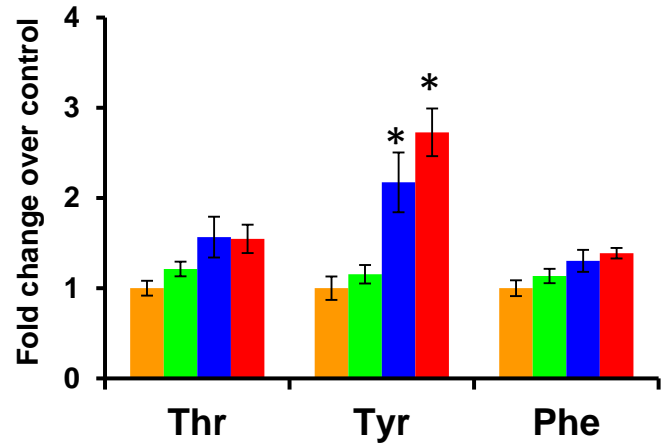
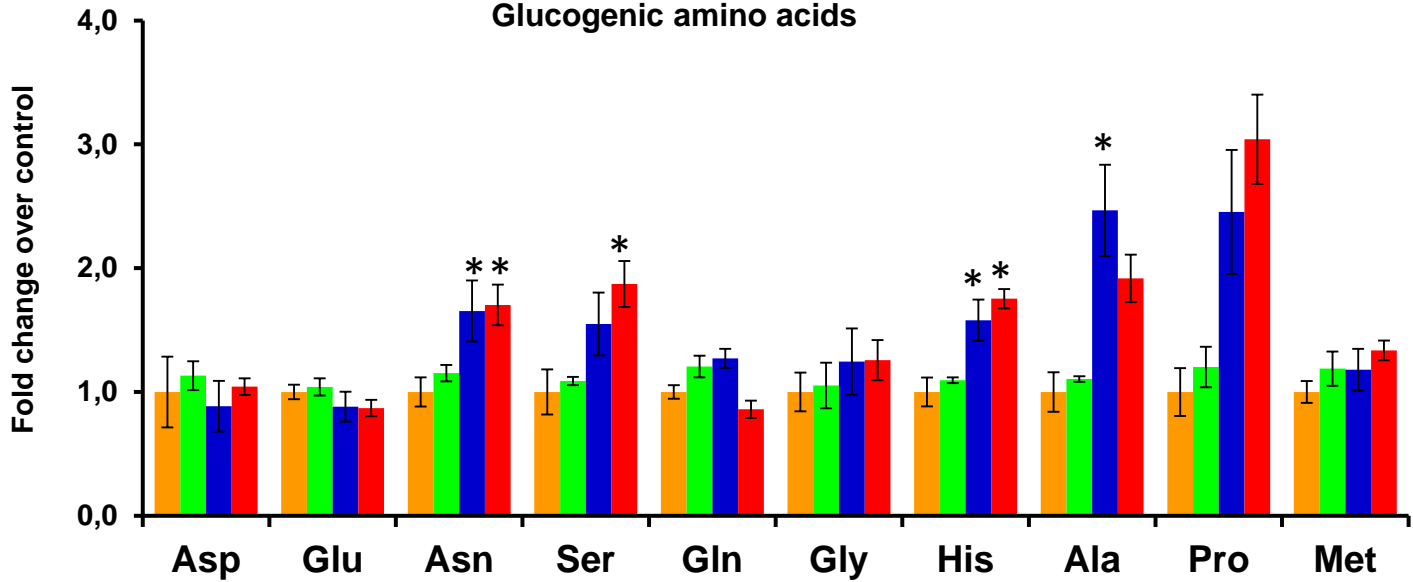
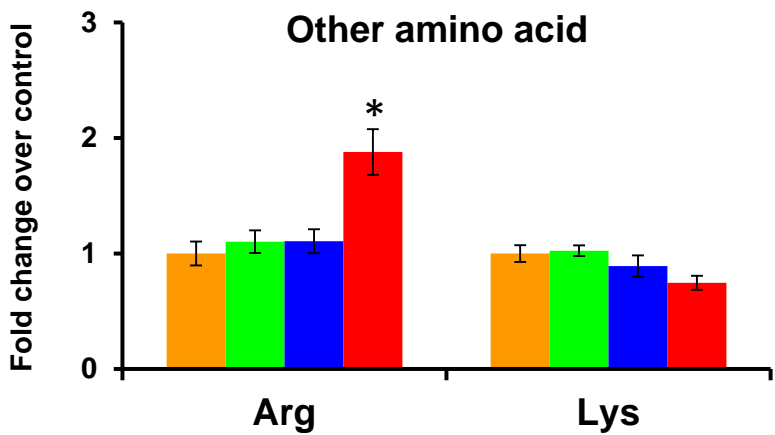
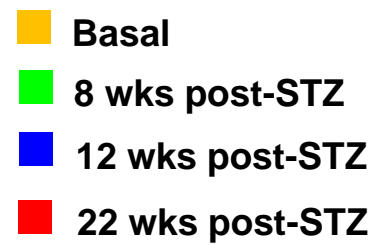

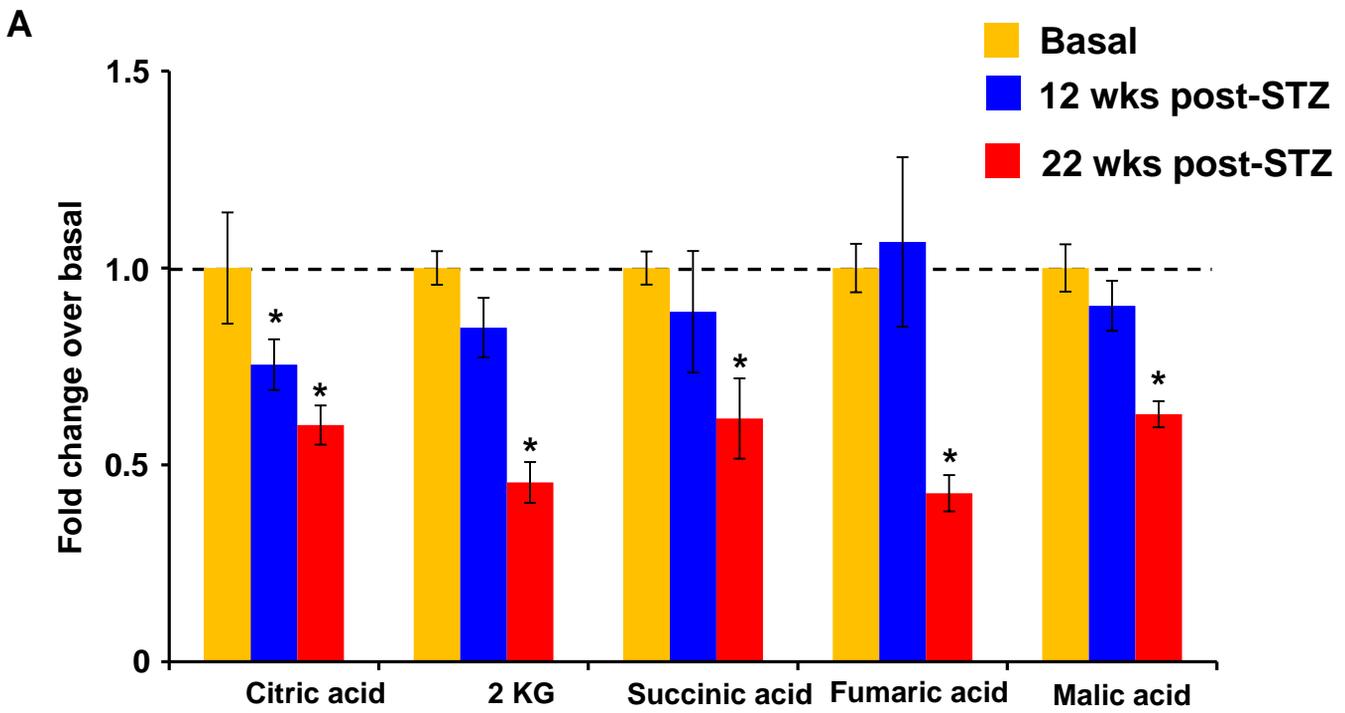
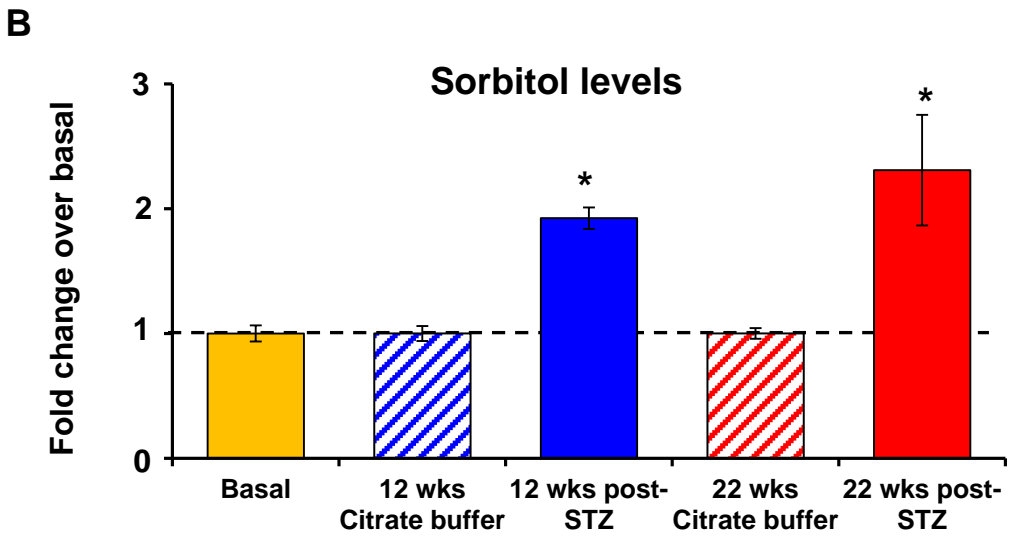
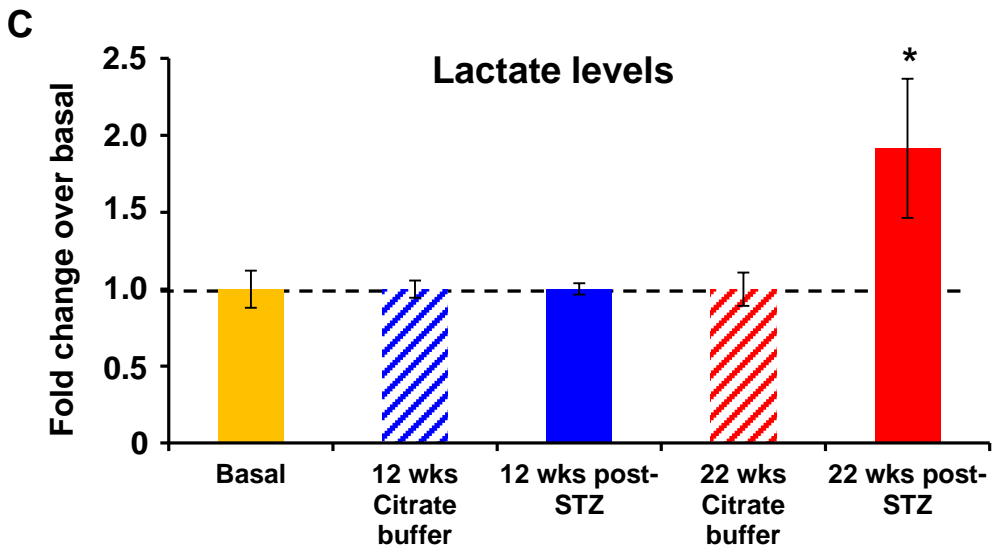